\documentclass[amsmath,amssymb,floatfix,showpacs,twocolumn,superscriptaddress]{revtex4}

\usepackage{graphicx}
\usepackage{dcolumn}
\usepackage{bm}

\begin{document}


\title{The Ehrenfest urn revisited: Playing the game on a realistic fluid model}

\author{Enrico Scalas}
\email{scalas@unipmn.it}
\homepage{www.econophysics.org, www.mfn.unipmn.it/~scalas}
\affiliation{Dipartimento di Scienze e Tecnologie Avanzate, Universit\`a del
Piemonte Orientale ``Amedeo Avogadro'', Via Bellini 25 G, 15100 Alessandria,
Italy}

\author{Edgar Martin}
\email{martine@staff.uni-marburg.de}
\affiliation{Fachbereich Chemie und WZMW, Philipps-Universit\"at Marburg,
35032 Marburg, Germany}

\author{Guido Germano}
\email{germano@staff.uni-marburg.de}
\homepage{www.staff.uni-marburg.de/~germano}
\affiliation{Fachbereich Chemie und WZMW, Philipps-Universit\"at Marburg,
35032 Marburg, Germany}

\date{May 9, 2007}



\begin{abstract}
The Ehrenfest urn process, also known as the dogs and fleas model, is
realistically simulated by molecular dynamics of the Lennard-Jones fluid.
The key variable is $\Delta z$, i.e.\ the absolute value of the difference
between the number of particles in one half of the simulation box and in the
other half. This is a pure-jump stochastic process induced, under coarse
graining, by the deterministic time evolution of the atomic coordinates.
We discuss the Markov hypothesis by analyzing the statistical properties of the
jumps and of the waiting times between the jumps. In the limit of a vanishing
integration time-step, the distribution of waiting times becomes closer to an
exponential and, therefore, the continuous-time jump stochastic process is
Markovian. The random variable $\Delta z$ behaves as a Markov chain and, in the
gas phase, the observed transition probabilities follow the predictions of the
Ehrenfest theory.
\end{abstract}

\maketitle

\section{Introduction}

A fundamental question in statistical mechanics is the reconciliation of the
irreversibility of thermodynamics with the reversibility of the microscopic
equations of motion governed by classical mechanics. In 1872 Ludwig Boltzmann
gave an answer with his $H$-theorem \cite{Boltzmann1872}, describing the
increase in the entropy of an ideal gas as an irreversible process. However,
the proof of this theorem contained the {\em Sto{\ss}zahlansatz}, i.e.\ the
assumption of molecular chaos. The result was subject to two main objections:
Loschmidt's {\em Umkehreinwand} (reversibility paradox) \cite{Loschmidt1876,
Loschmidt1877} and Zermelo's {\em Wiederkehreinwand} (recurrence paradox)
\cite{Zermelo1896}. Boltzmann's reply to the two objections was not fully
understood at the time, but is now considered as a corner-stone of statistical
mechanics. It is summarized in the article that Paul and Tatiana Ehrenfest
wrote for the German {\em Encyclopedia of Mathematical Sciences}
\cite{Ehrenfest1911}. Subsequently, Boltzmann's approach has been reformulated
in the language of stochastic processes \cite{terHaar1954,terHaar1955,
Penrose1970}.

Essentially, even in the presence of a deterministic microscopic dynamics,
the {\em coarse graining} of configuration space due to the observer's state of
knowledge results in a stochastic process, where the number of particles in a
given cell varies at random as a function of time.


Exactly 100 years ago \cite{Ehrenfest1907}, the Ehrenfests gave a simple and
convincing interpretation of Boltzmann's ideas in term of an urn stochastic
process that is a periodic Markov chain in their original formulation
\cite{Ehrenfest1911,Hoel1972,Costantini2004}. There are $N$ marbles or balls
to be divided into two equal parts of a box. In order to fix the ideas, let us
call $P$ the number of balls in the left part and $Q$ the number of balls in
the right part. The balls are labeled from 1 to $N$. At each step of the
process, an integer between 1 and $N$ is selected with probability $1/N$ and
the corresponding ball is moved from one part to the other. Rather than urns
and balls, later variants of the model used dogs and fleas jumping from one
dog to the other, but this does not change the mathematics. Indeed, according
to Ref.~\onlinecite{Costantini2004}, the Ehrenfests already had something
similar to fleas in mind because they used the verb {\em h\"upfen}, meaning
{\em hop}, that is more appropiate for fleas than for marbles. Assuming $P>Q$,
in terms of the random variable $\Delta z = |P-Q|$, the unconditional
equilibrium probability of a certain value of $\Delta z$ is given by
\begin{equation}
\label{unconditional}
p_\mathrm{eq}(\Delta z) = \binom{N}{P} \left( \frac{1}{2} \right)^{N} =
\binom{N}{(N+\Delta z)/2} \left( \frac{1}{2} \right)^{N}.
\end{equation}
In the limit for $N \to \infty$ \cite{terHaar1954}
\begin{equation}
\label{uncond_limit}
p_\mathrm{eq}(\Delta z)
\sim \sqrt{\frac{2\pi}{N}}\exp\left(-\frac{(\Delta z)^2}{2N}\right).
\end{equation}
The transition probabilities of a decrease, $p_\mathrm{d}(\Delta z - 2\,|\,
\Delta z)$, and of an increase, $p_\mathrm{u}(\Delta z + 2\,|\,\Delta z)$,
of $\Delta z$ are given by
\begin{subequations}
\label{transition}
\begin{eqnarray}
p_\mathrm{d}(\Delta z-2\,|\,\Delta z) &=& \frac{P}{N} = \frac{N+\Delta z}{2N}\\
p_\mathrm{u}(\Delta z+2\,|\,\Delta z) &=& \frac{Q}{N} = \frac{N-\Delta z}{2N}.
\end{eqnarray}
\end{subequations}
Eqs.~(\ref{transition}) completely determine the Ehrenfest urn Markov chain.
It is possible to define an aperiodic version of this process, but both
versions share the same stationary distribution (invariant measure) given by
Eq.~(\ref{unconditional}), that in the aperiodic case is also the equilibrium
distribution \cite{Hoel1972,Sinai1992}. As noticed by Kohlrausch and
Schr\"odinger \cite{Kohlrausch1926,Godreche2002}, Eq.~(\ref{unconditional}) can
be regarded as the equilibrium distribution for a fictitious walker obeying a
suitable forward Kolmogorov equation:
\begin{equation}
\label{forward_Kolmogorov}
p(P,t+1) = \frac{P+1}{N}p(P+1,t) + \frac{N-P+1}{N}p(P-1,t).
\end{equation}
By means of this stochastic process, the Ehrenfests were able to present
convincing evidence in favour of Boltzmann's approach. In this example, the
random variable $\Delta z$ is the analogous of $H$ and it almost always
decreases from any higher value; moreover this is true both in the direct and
reverse time direction as required by Loschmidt's {\em Umkehreinwand}, and
$\Delta z$ is quasiperiodic as required by Zermelo's {\em Wiederkehreinwand}
\cite{Ehrenfest1911}.

But what happens if this game is played with a real fluid or, more modestly,
with a realistic model \cite{Rahman1964,Verlet1967} of a fluid? As argued by
Boltzmann, in this case the deterministic microscopic dynamics induces a
stochastic process and, again, the number of fluid particles in the left side
of the box $P$ and in the right side of the box $Q$ fluctuate as a function of
time. Here, the coarse graining is simply due to the division into two equal
parts of the box that contains the fluid. The Markov hypothesis, clearly
explained by Penrose \cite{Penrose1970}, is instrumental in deriving the
properties of statistical equilibrium. There is, however, a further
complication. $P$, $Q$, and $\Delta z$ can be constant for a certain time
before changing their values. The waiting times between these jumps are
randomly distributed as well. The mathematical model for such a process is
called a {\em continous-time pure-jump stochastic process} \cite{Hoel1972}.
A pure-jump process is Markovian if and only if the waiting time between two
consecutive jumps is exponentially distributed (this distribution may depend on
the initial non-absorbing state) \cite{Hoel1972}. The following remark is
important. It is possible to define a pure-jump process by coupling a Markov
chain, such as the Ehrenfest urn process defined above, with a point process
for the inter-jump waiting times. If the latter is non exponential, the
pure-jump process is non-Markovian.

In the present work, we investigate the Markovian character of the pure-jump
process induced by the simulation of a Lennard-Jones fluid in a box. 

\section{Methodology}
\label{sec:methodology}

\begin{figure}[h]
\includegraphics[angle=-90,width=\columnwidth]{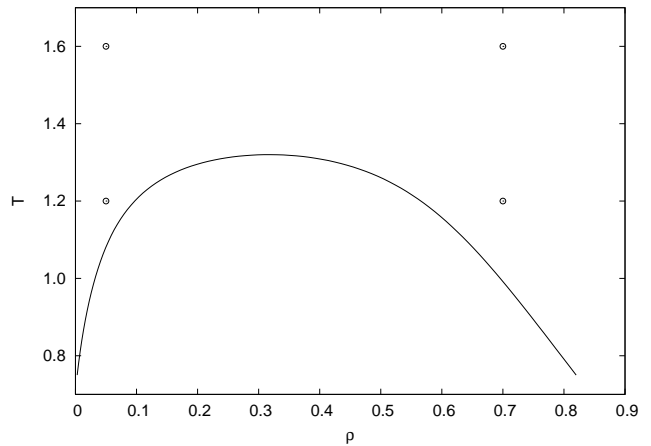}
\caption{\label{fig:phase_diagram}
The four simulated points (circles) in the phase diagram of the Lennard-Jones
fluid. The liquid-vapour curve (solid line) is a Bezier fit to data from
Ref.~\onlinecite{Panagiotopoulos1987}. The critical point corresponds to the
maximum of the liquid-vapour curve.}
\end{figure}

Systems with $N = 500,\ 1000,\ 2000\ \textrm{and}\ 100\,000$ atoms interacting
with the cut and shifted Lennard-Jones pair potential
\begin{eqnarray}
U &=& \sum_{i<j} \left[ U_{ij} (r_{ij}) - U_{ij} (r_\textrm{cut})\right], \\
U_{ij}(r_{ij}) &=& 4\epsilon \left[ \left( \frac{\sigma}{r_{ij}} \right)^{12}
-\left( \frac{\sigma}{r_{ij}} \right)^6 \right], \nonumber
\end{eqnarray}
where $r_{ij}$ is the interatomic distance, were simulated using classical
molecular dynamics \cite{Allen-Tildesley1989,Frenkel-Smit2002}. We employed a
parallelepiped unit box with side ratios 1:1:1 when $N = 1000$ or 2:1:1 in the
other cases, and periodic boundary conditions in all three directions of space.
For $N = 1000$, we used also two parallel soft walls in the $x$-direction with
periodic boundary conditions in the $y,z$-directions only, i.e.\ ``slab''
boundary conditions. The wall potential was given by integrating the
Lennard-Jones potential over a semi-infinite wall of atoms distributed with a
density $\rho_{\textrm{w}}$ \cite{Steele1973}:
\begin{eqnarray}
U_\textrm{w} &=& \sum_i \left[U_{i\textrm{w}} (r_{i\textrm{w}}) -
U_{i\textrm{w}} (r^\textrm{w}_\textrm{cut}) \right],\\
U_{i\textrm{w}}(r_{i\textrm{w}}) &=&
4\pi\rho_\textrm{w} \sigma^3\epsilon\left[
\frac{1}{45}\left(\frac{\sigma}{r_{i\textrm{w}}}\right)^{9}
-\frac{1}{6}\left(\frac{\sigma}{r_{i\textrm{w}}}\right)^3\right], \nonumber
\end{eqnarray}
where $r_{i\textrm{w}}$ is the atom-wall distance. We did not put walls along
all three directions of space to avoid too large surface effects with small
values of $N$. We use reduced units with $\sigma = \epsilon = m = k_\textrm{B}
= 1$, where $m$ is the mass of each atom and $k_\textrm{B}$ is the Boltzmann
constant. This defines the time unit as $\sigma\sqrt{m/\epsilon}$ and the
temperature unit as $\epsilon/k_\textrm{B}$.
We used the common bulk cutoff value $r_\textrm{cut} = 2.7$ and a wall cutoff
$r^\textrm{w}_\textrm{cut} = \sqrt[6]{2/5}$ corresponding to the minimum of
the wall potential, so that the cut and shifted wall potential is purely
repulsive. $\rho_\textrm{w}$ was set to 1, i.e.\ slightly below the densities
of bcc (1.06) and fcc (1.09) lattices.
We chose four points in the phase diagram with $(\rho,\,T) = (0.05, 1.2),\
(0.7, 1.2),\ (0.05, 1.6),\ (0.7, 1.6)$ lying around the critical point, whose
accepted value for the Lennard-Jones fluid is (0.35,1.35) \cite{Nicolas1979,
Panagiotopoulos1987}; see Fig.~\ref{fig:phase_diagram}.

\begin{figure}[h]
\includegraphics[angle=-90,width=\columnwidth]{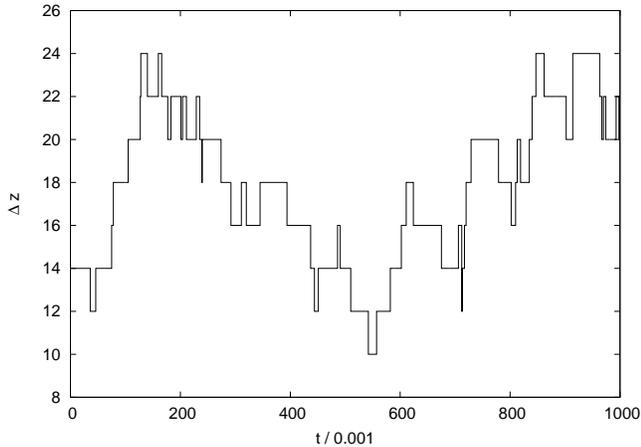}
\caption{\label{fig:delta_z}
The pure-jump stochastic process $\Delta z = |P-Q|$ as a function of the first
1000 time steps of the first simulation run in Table~\ref{tab:1}.}
\end{figure}

Production runs of 10 million time steps were done in the microcanonic ensemble
with the velocity Verlet integrator \cite{Swope1982,Tuckerman1992}, while
equilibration runs were performed in the canonic ensemble with an extended
system thermostat \cite{Tuckerman1992,Nose1984,Hoover1985,Sergi1999}.
At every time-step we measured $P$ as the number of atoms on the left part of
the box, that is with $r_x < 0$. Thus, as mentioned before, one has $\Delta z =
|P-Q| = |2P-N|$; see Fig.~\ref{fig:delta_z}. While a time-step $\Delta t =
0.025$ is sufficient for an acceptable energy conservation in this kind of
system \cite{Sergi1999}, to get a good resolution of the waiting times we
started employing a smaller $\Delta t = 0.001$; for $N = 1000$, we obtained
$\sigma_E / |\langle E \rangle|$ in the range from $7.0\cdot10^{-6}$ to
$1.1\cdot10^{-4}$ depending on $\rho$ and $T$.
Nevertheless, any time-step we tried down to 0.0001 was still large enough to
observe a few percent of jumps in $\Delta z$ greater than 2; the shorter the
average waiting time, the higher the percentage. There were even occasional
variations greater than 4 or, for some parameter combinations, 6, 8 or 10.

A trajectory of 10 million time-steps with $N = 1000$ took about 20 hours at
$\rho = 0.05$ and about 80 hours at $\rho = 0.7$ on a 2.4 GHz Intel Pentium
IV processor with our own C++ code using Verlet neighbour lists. With
$N = 100\,000$, the lower density lasted 17.5 hours on 64 IBM Power4+
processors at 1.7 GHz, and the higher density almost 9 days on 64 AMD Opteron
270 processors at 2.0 GHz, with a Fortran code using domain decomposition and
linked cell lists \cite{Wilson1997}. Trajectories of this length are the main
difference with respect to the pioneering simulations of 40 years ago, when for
$N = 864$ atoms and $\rho \simeq 0.8$ one time-step took 45 seconds on a
CDC-3600 \cite{Rahman1964}, while trajectories consisted typically of 1200
time-steps \cite{Verlet1967}.

\section{Results}

\subsection{Analysis of jumps}

In this section, we study the random variable $\Delta z$. We compare simulation
results with the Ehrenfest theory to see whether $\Delta z$ obeys the
Markov-chain equations (\ref{unconditional}--\ref{forward_Kolmogorov}).

\begin{figure}[h]
\includegraphics[angle=-90,width=\columnwidth]{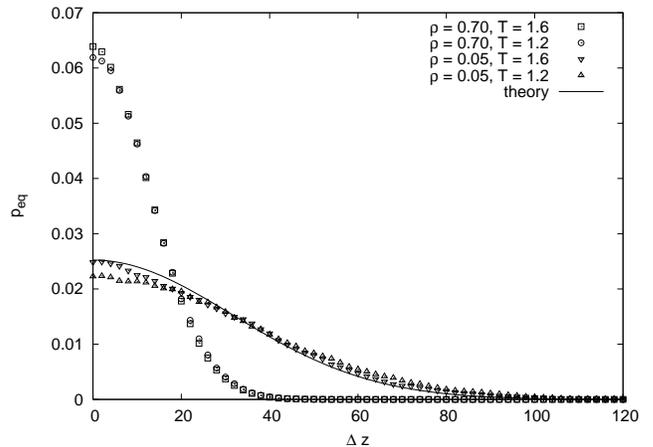}
\caption{\label{fig:delta_zh}
Histograms of the values of $\Delta z$ from the runs of the $N = 1000$ systems
without walls. The theoretical line given by Eq.~(\ref{uncond_limit}) matches
the gas states.}
\end{figure}

In Fig.~\ref{fig:delta_zh}, the empirical estimate for $p_\mathrm{eq} (\Delta
z)$ is plotted and compared with Eq.~(\ref{uncond_limit}). There is visibly a
good agreement between the quantitative predition of Eq.~(\ref{uncond_limit})
and the empirical histogram for the gas phase, and this agreement is slightly
better for the higher temperature.

In Fig.~\ref{fig:delta_z1tp}, we report results on the one-step transition
probabilities. The Ehrenfest prediction is given by Eqs.~({\ref{transition}).
Again, in the gas phase of the Lennard-Jones fluid there is agreement between
the sampled transition probabilities and the Ehrenfest theory. Even if linear
in $\Delta z$, the sampled transition probabilities for the liquid phase
deviate from Eqs.~({\ref{transition}).

\begin{figure}[h]
\includegraphics[angle=-90,width=\columnwidth]{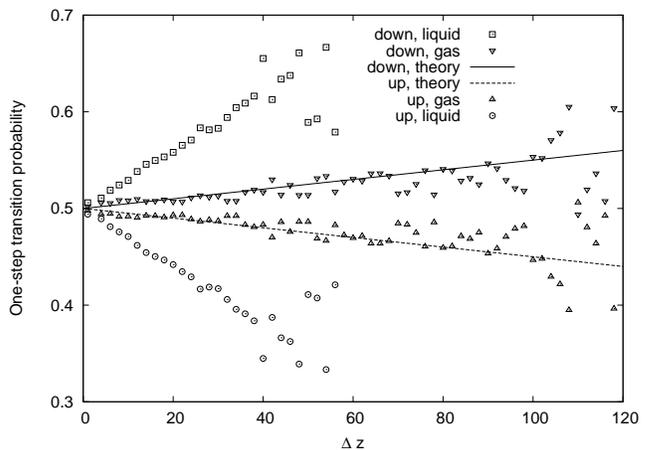}
\caption{\label{fig:delta_z1tp}
One-step transition probabilities $p_\mathrm{d}(\Delta z - 2\,|\,\Delta z)$ and
$p_\mathrm{u}(\Delta z + 2\,|\,\Delta z)$ for $\rho = 0.7,\ T = 1.2$ (liquid)
and $\rho = 0.05, T = 1.6$ (gas), $N = 1000$ without walls. The theoretical
lines $1/2\pm\Delta z/(2N)$ \cite{terHaar1954} 
match the gas state.}
\end{figure}

Sampled two-steps transition probabilites are plotted in
Fig.~\ref{fig:delta_z2tp}. If the process is a Markov chain, these
probabilities must be the product of two one-step transition probabilities.
This property appears satisfied both for the gas and for the liquid.
Moreover, for the gas, the sampled two-steps probabilities follow the Ehrenfest
quantitative prediction given by Eqs.~({\ref{transition}).

\begin{figure}[h]
\includegraphics[angle=-90,width=\columnwidth]{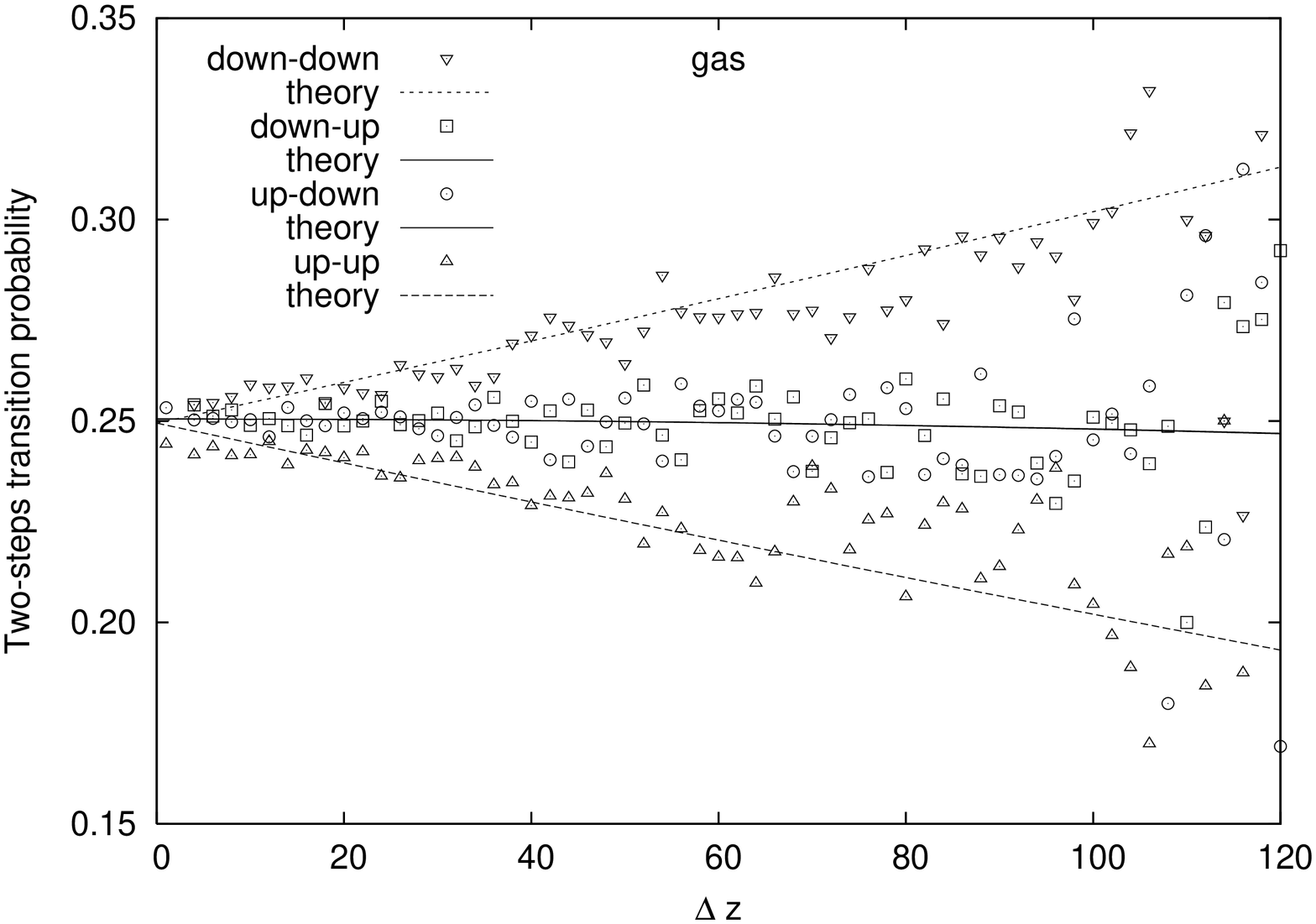}\\
\includegraphics[angle=-90,width=\columnwidth]{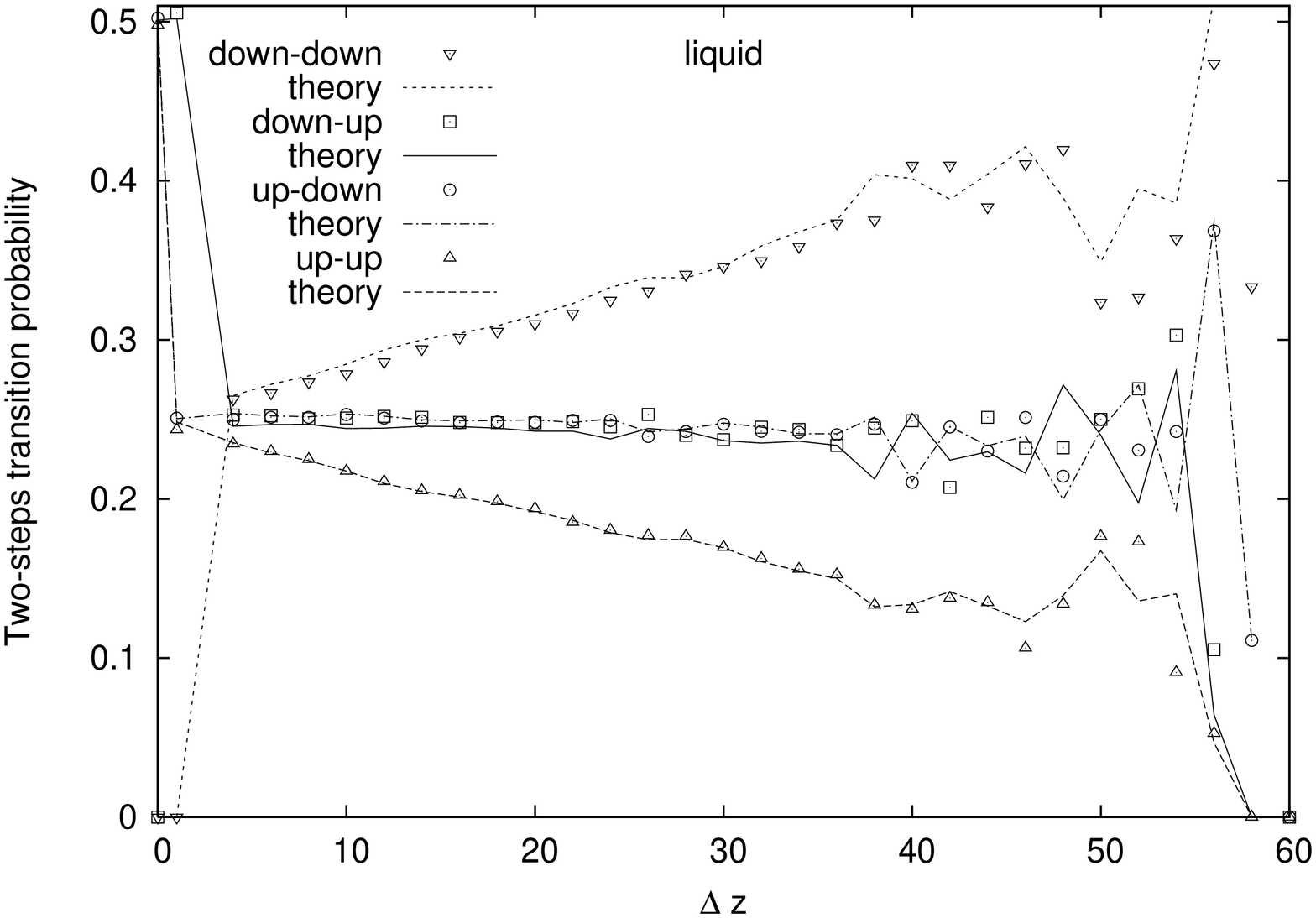}
\caption{\label{fig:delta_z2tp}
Two-steps transition probabilities
$p_\mathrm{dd}(\Delta z - 4\,|\,\Delta z)$,
$p_\mathrm{du}(\Delta z    \,|\,\Delta z)$,
$p_\mathrm{ud}(\Delta z    \,|\,\Delta z)$ and
$p_\mathrm{uu}(\Delta z + 4\,|\,\Delta z)$ for
$\rho = 0.05,\ T = 1.6$ (gas, top) and $\rho = 0.7,\ T = 1.2$ (liquid, bottom),
$N = 1000$ without walls. The theoretical lines are the product of the two
corresponding one-step transition probabilities, e.g.\
$p_\mathrm{uu}(\Delta z + 4\,|\,\Delta z) =
p_\mathrm{u}(\Delta z + 4\,|\,\Delta z + 2)
p_\mathrm{u}(\Delta z + 2\,|\,\Delta z)$. We use the theoretical one-step
transition probabilities for the gas and the observed ones for the liquid.}
\end{figure}

Even if, rigorously speaking, we have not shown that, for all $n$, the $n$-step
transition probabilities are the product of $n$ one-step transition
probabilities (see Ref.~\onlinecite{Feller1959} for processes obeying the
semigroup property that are not Markov chains), at least we can claim that we
have not been able to falsify the Markov-chain hypothesis for $\Delta z$ based
on our statistics in all the investigated cases. Remarkably, the pure Ehrenfest
Markov-chain theory is a good approximation for the gas, but does not work for
the liquid.

\subsection{Analysis of waiting times}

\begin{table}[h]
\begin{center}
\begin{ruledtabular}
\begin{tabular}{rrrr|rrrr}
$\Delta t$ & $N$ & $\rho$ & $T$ &
$n$ & $A^2$ & $\langle \tau \rangle $ & $\sigma_\tau$\\
\hline
1.0&    1000   &.05&1.2&$   613\,751$&     2061&16.29&15.79\\
1.0&  w 1000   &.05&1.2&$   618\,220$&     2096&16.18&15.69\\
1.0&    1000   &.05&1.6&$   704\,881$&     3038&14.19&13.67\\
1.0&  w 1000   &.05&1.6&$   704\,007$&     3031&14.20&13.68\\
1.0&    1000   &.70&1.2&$1\,386\,970$&$18\,666$&7.210&6.662\\
1.0&  w 1000   &.70&1.2&$1\,407\,654$&$19\,428$&7.104&6.562\\
1.0&    1000   &.70&1.6&$1\,578\,866$&$26\,525$&6.334&5.779\\
1.0&  w 1000   &.70&1.6&$1\,565\,301$&$25\,835$&6.389&5.841\\
1.0&     500   &.70&1.6&$   675\,876$&     2847&14.80&14.14\\
1.0&    2000   &.70&1.6&$1\,561\,554$&$25\,704$&6.404&5.856\\
0.2&    2000   &.05&1.2&$   127\,237$&    29.84&15.72&15.59\\
0.1&    2000   &.05&1.2&$    64\,617$&     3.78&15.48&15.46\\
.01&    2000   &.05&1.2&$     6\,306$&    0.686&15.85&16.15\\
1.0&$100\,000$ &.05&1.2&$4\,988\,531$&$587\,570$&2.005&1.419\\
0.1&$100\,000$ &.05&1.2&$   820\,837$&     4534 &1.218&1.166\\
0.1&$100\,000$ &.70&1.6&$2\,043\,142$&$ 52\,278$&.4894&.4369  
\end{tabular}
\end{ruledtabular}
\end{center}
\caption{\label{tab:1}
For each integration time-step $\Delta t$, number of atoms $N$, density $\rho$
and temperature $T$ (a ``w'' before the $N$ value indicates a system with walls
in the $x$-direction), this table gives the number of observed waiting times
$n$, the values of the Anderson-Darling statistics $A^{2}$ \cite{Stephens1974},
the average waiting time $\langle \tau \rangle$, and the standard deviation of
waiting times $\sigma_\tau$. Reduced units as defined in
Sec.~\ref{sec:methodology} are used throughout, with times divided by 0.001.
The standard error on $\langle \tau \rangle$ is around 0.02 for $\rho = 0.05$
and 0.006 for $\rho = 0.70$. The standard error on $\sigma_\tau$ is around 0.02
for $\rho = 0.05$ and 0.005 for $\rho = 0.70$. Only significant digits are
given in the table. The last digit of $\langle \tau \rangle$ and $\sigma_\tau$
is of the same order of magnitude as $\sigma_\tau/\sqrt{n}$. See text for
further explanations.}
\end{table}

\begin{figure}[h]
\includegraphics[angle=-90,width=\columnwidth]{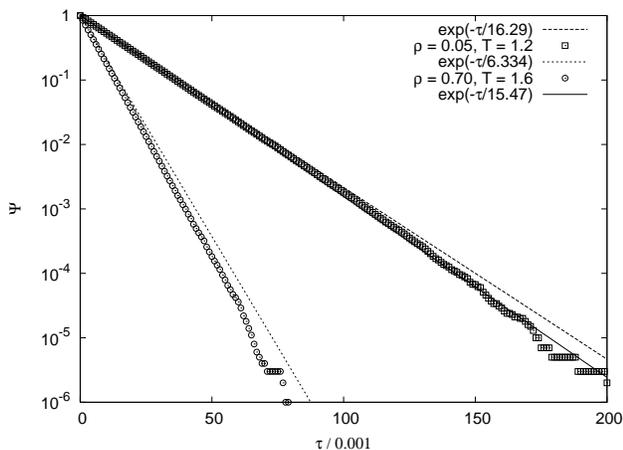}
\caption{\label{fig:survival_function_1-7}
$N = 1000$, no walls. Comparison between the observed survival functions and
the theoretical exponential survival functions (dashed lines) with the
corresponding average waiting time $\langle \tau \rangle$, for the closest
case (squares) and the most distant case (circles). The theoretical exponential
survival function of the system with $N = 2000$ and $\Delta t = 10^{-5}$ is
shown for reference (continuous line).}
\end{figure}

The results of the simulations regarding the waiting time distribution are
summarized in Table~\ref{tab:1}.
The Anderson-Darling statistics $A^2$ reported in the sixth column results from
\cite{Stephens1974}
\begin{eqnarray}
\label{asquared}
A^2 & = & \left\{ -\sum_{i=1}^{n} \frac{(2i-1)}{n} [\ln \Psi(\tau_{n+1-i})
\right. \\ & & \left. \phantom{\sum_{i=1}^{n}} + \ln (1-\Psi(\tau_i))] -n
\right\} \left(1+\frac{0.6}{n} \right), \nonumber
\end{eqnarray}
where $\Psi(\tau)$ denotes the survival function, a short name for the
complementary cumulative distribution function, i.e.\ the probability that
waiting times are larger than $\tau$. In Eq.~(\ref{asquared}) the waiting times
are sorted: $\tau_1 \leq \tau_2 \leq \ldots \leq \tau_n$. The limiting value at
$1\%$ significance for accepting the null hypothesis of exponentially
distributed waiting times is 1.957. Therefore, the null hypothesis can be
rejected in all cases with $\Delta t \geq 0.0001$. The average waiting time
$\langle \tau \rangle$ and the standard deviation $\sigma_\tau$ of the observed
distribution, reported in columns seven and eight, must coincide for an
exponential distribution. Even if their values are close, with the given
statistics they cannot be considered equal.
Fig.~\ref{fig:survival_function_1-7} further illustrates this point;
there, the {\em closest} case to an exponential for $N = 1000$ is presented,
$\rho =0.05,\ T = 1.2$ without walls, as well as the {\em most distant} case,
$\rho = 0.7,\ T = 1.6$ without walls. In both cases the points are the observed
survival function, $\Psi(\tau)$, and the dashed line is the exponential fit.
A deviation from the exponential distribution is evident at first sight. It is
important to remark that this is a one-parameter fit, since the average waiting
time $\langle \tau \rangle$ is sufficient to fully determine the exponential
distribution, with survival function $\Psi_{\textrm{exp}}(\tau) = \exp (-\tau /
\langle \tau \rangle)$, corresponding to a given data set. In other words, the
mere fact that in log-linear scale the survival function is approximately a
straight line is not sufficient to conclude that the observed distribution is
exponential. In the four cases studied here, the presence of walls does not
significantly affect the results.

However, the agreement improves if the integration time-step $\Delta t$ is
reduced from 0.001 to 0.0002: for $\rho = 0.05,\ T = 1.2$ in the $N = 2000$
system, $A^2$ drops from 2061 to 29.84 and $\langle \tau \rangle$ from 16.29 to
15.72; the lower value of $\langle \tau \rangle$ corresponds better to the
observed survival function. The data change very little with respect to $\Delta
t = 0.001$ and are not shown in Fig.~\ref{fig:survival_function_1-7} to avoid
cluttering. This indicates that the discrepancy is due to the finite
integration time-step and can be controlled through the latter. The hypothesis
is confirmed reducing $\Delta t$ further: for $\Delta t = 10^{-4},\ A^2 =3.78$,
and finally for $\Delta t = 10^{-5},\ A^2 = 0.686 < 1.957$, i.e.\ the required
threshold. The same trend is evident in the $N = 100\,000$ system, see
Fig.~\ref{fig:survival_function_12-13}, though even smaller time steps would be
necessary to reach the threshold because the average waiting time decreasess
inversely proportionally to the interface area.

\begin{figure}[h]
\includegraphics[angle=-90,width=\columnwidth]{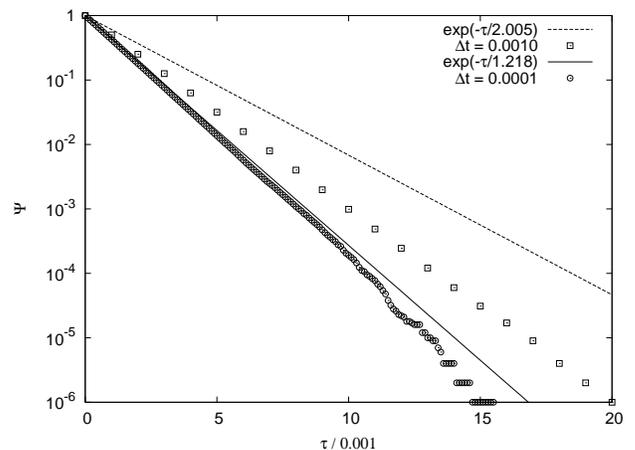}
\caption{\label{fig:survival_function_12-13}
Reducing the integration time-step $\Delta t$ improves the agreement between
the observed survival function and an exponential function with a time constant
equal to the average waiting time; system with $N = 100\,000,\ \rho = 0.05,\
T = 1.2$.}
\end{figure}

As suggested by intuition, the average waiting time decreases with higher
density and temperature, but also whith a larger interface area $S$ between the
two parts of the box. Actually, the product $\langle \tau \rangle S$ is a
constant for a given density and temperature. The survival functions of systems
with different sizes overlap if $\langle \tau \rangle$ is multiplied with the
interface area. This is shown in Fig.~\ref{fig:surface}, where it is also clear
that there are no changes due to the finite size of the system for $N \geq
1000$ (after correcting for the interface area, the survival function of $N =
500$ is slightly displaced from all the others).

\begin{figure}[h]
\includegraphics[angle=-90,width=\columnwidth]{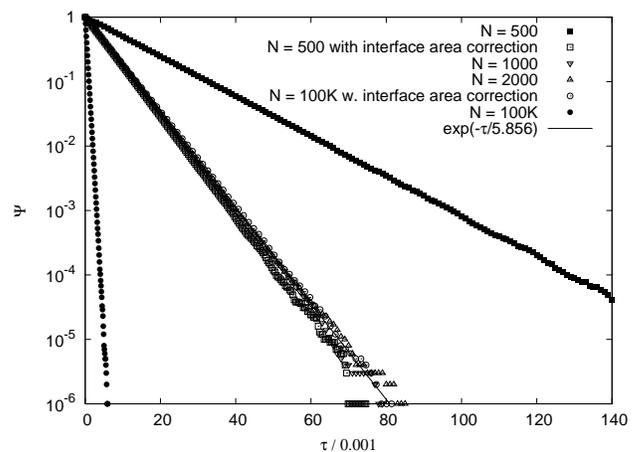}
\caption{\label{fig:surface}
Survival functions for $\rho = 0.7,\ T 1.6)$ and different system sizes.
They overlap if $\langle \tau \rangle$ is multiplied with the ratio of the
interface area to the interface area of the systems with $N = 1000$ or 2000
(that are equal because the former is the only one with a cubic unit box, while
all the others have side ratios of 2:1:1). A finite-size effect is noticeable
only on the smallest system.}
\end{figure}

A better strategy than reducing the time-step is to interpolate the time of the
barrier crossing within a conventional time-step: this way the waiting times
can be determined with floating-point precision rather than as integer
multiples of $\Delta t$, there will not be changes in $\Delta z > 2$, and it is
likely that good results can be obtained with the maximum $\Delta t$ compatible
with energy conservation. Though we believe that the major effect of a finite
$\Delta t$ is through sampling, because without interpolation waiting times are
systematically overestimated by a fraction of $\Delta t$, another effect is
through the approximation of the true canonical dynamics. Indeed, with a soft
potential this approximation can be reduced only in the limit of $\Delta t \to
0$, but it can be avoided completely in a system of hard spheres. Work on both
lines, interpolation of the waiting times and hard spheres, is in progress.

\section{Conclusions}

In summary, we have studied the Ehrenfest urn whith a realistic model of
condensed matter, the Lennard-Jones fluid. The Ehrenfest urn has been defined
by Mark Kac the best model ever envisaged in statistical mechanics
\cite{Kac1959}, yet it has also been criticized as a marvellous exercise too
far removed from reality \cite{Costantini2004}. In the 100th anniversary of the
Ehrenfests' original paper, we have shown that this criticism is unjustified,
since computer ``experiments'' allow to follow the motion of molecules and to
count how many are on one side of a box or the other at a given time. We have
studied the behaviour of the pure-jump stochastic process $\Delta z = |P-Q|$
induced by the deterministic dynamics under coarse graining, where $P$ is the
number of fluid particles on the left-hand side of the simulation box and $Q$
that on the right-hand side. We have performed simulations with periodic
boundary conditions and with walls in one direction, finding that the presence
of walls does not affect the results. We have found that in the gas phase the
observed transition probabilities follow the predictions of the Ehrenfest
theory, and that the waiting time distribution between successive variations
of $\Delta z$, though not strictly exponential, becomes closer to an
exponential reducing the integration time-step; therefore, in the limit of a
vanishing time-step, we found that the corresponding pure-jump process is
Markovian. To our knowledge, this is the first characterization of a pure-jump
stochastic process induced by a deterministic dynamics under coarse-graining.
In the future, we plan to further study the stochastic process presented here
interpolating the waiting times to higher precision, simulating systems of
hard spheres to avoid approximations in the dynamics due to a finite
integration time-step, and investigating the pure-jump process in a
coarse-grained configuration space as required by the theory developed by
Boltzmann. Our results so far corroborate the Markovian hypothesis lying at the
foundation of statistical mechanics \cite{Penrose1970}.

\section*{Acknowledgements}
E.~Scalas is grateful for support from the Philipps-University Marburg that
sponsored a short visit during which this paper was conceived. The systems with
$N = 100\,000$ and $\rho = 0.05$ were run on an IBM Regatta p690+ parallel
computer at the John von Neumann Institute for Computing of the
Forschungszentrum J\"ulich.

\end{document}